\documentclass[preprint]{aastex}
\usepackage{graphicx}
\usepackage{bm}
\usepackage{epsfig}
\usepackage{graphicx}
\usepackage{subfigure}
\usepackage{amsmath}
\usepackage{natbib}
\usepackage{multirow}

\shorttitle{The MHD model of kHz QPOs}

\shortauthors{C. Shi et al.}

\begin{document}


\title{The Magneto Hydro Dynamical Model of KHz Quasi Periodic Oscillations in Neutron Star Low Mass X-ray Binaries (\uppercase \expandafter {\romannumeral 2})}

\author{Chang-Sheng Shi\altaffilmark{1,4,6}, Shuang-Nan Zhang\altaffilmark{1,2,5 \star}, Xiang-Dong Li\altaffilmark{3,6}}

\altaffiltext{1}{National Astronomical Observatories, Chinese Academy of Sciences, Beijing 100012, China}
\altaffiltext{2}{Key Laboratory of Particle Astrophysics, Institute of High Energy Physics,
Chinese Academy of Sciences, Beijing 100049, China; zhangsn@ihep.ac.cn}
\altaffiltext{3}{Department of Astronomy, Nanjing University, Nanjing 210093, China}
\altaffiltext{4}{College of Material Science and Chemical Engineering,
Hainan University, Hainan 570228, China}
\altaffiltext{5}{Physics Department, University of Alabama in
Huntsville, Huntsville, AL 35899, USA}
\altaffiltext{6}{Key Laboratory of
Modern Astronomy and Astrophysics (Nanjing University), Ministry of
Education, Nanjing 210093, China}
\email{$\star$ zhangsn@ihep.ac.cn}

\begin{abstract}
We study the kilohertz quasi-periodic oscillations (kHz QPOs) in neutron star low mass X-ray binaries (LMXBs) with a new magnetohydrodynamics (MHD) model, in which the compressed magnetosphere is considered. The previous MHD model (Shi \& Li 2009) is re-examined and the relation between the frequencies of the kHz QPOs and the accretion rate in LMXBs is obtained. Our result agrees with the observations of six sources (4U 0614+09, 4U 1636--53, 4U 1608--52, 4U 1915--15, 4U 1728--34, XTE 1807--294) with measured spins. In this model the kHz QPOs originate from the MHD waves in the compressed magnetosphere. The single kHz QPOs and twin kHz QPOs are produced in two different parts of the accretion disk and the boundary is close to the corotation radius. The lower QPO frequency in a frequency-accretion rate diagram is cut off at low accretion rate and the twin kHz QPOs encounter a top ceiling at high accretion rate due to the restriction of innermost stable circular orbit.
\end{abstract}


\section{Introduction}

The fastest variability, i.e., the high frequency quasi-periodic
oscillations (QPOs), has been observed in both neutron star low mass X-ray binaries (NS-LMXBs) and black hole LMXBs, which gives us an important channel to understand the physics
of accretion process in the accretion disks of LMXBs.
Especially it gives us a clue to find the parameters of the compact stars, e.g. their equation of state (van der Klis et al. 2006). In NS-LMXBs the frequencies of the high frequency QPOs approach the Kepler frequency of the matter at the surface of the NS
and often exceed 1000 Hz, so named as kilohertz (kHz) QPOs. The kHz QPOs always appear as broad peaks in their Fourier power spectra and were often discovered in pairs; they are labeled as upper QPOs ($\nu_{\rm upper}$) and lower QPOs ($\nu_{\rm lower}$) according to their frequencies. The kHz QPOs change with the source states and the X-ray count rates (M\'endez et al. 1999) (probably the mass accretion rate). In addition there is an interesting phenomenon that the X-ray intensity and kHz QPOs do not exhibit a one-to-one relation in 4U 1608--52 and 4U 1636--53, i.e. ``the parallel track" (M\'endez 2003). Barret et al. (2005, 2006) discovered that the lower QPO frequency from 4U 1636--53 in the frequency-count rate diagram has a maximum value (around 920 Hz), above which frequency no QPO was detected for any count rate; they named this phenomenon the ceiling of the frequency. In this work, we simply call this maximum value the  ceiling frequency.

Many models have been suggested to explain the physics of kHz QPOs in NS-LMXBs
(see e.g. van der klis 2006, Shi \& Li 2009).
Lin et al. (2011) discussed the frequency relationship of kHz QPOs for 4U 1636--53
and Sco X--1 by comparing the observations to some theoretical models. Generally the models of
the kHz QPOs mainly include several types, i.e. beat-frequency models (Miller et al. 1998; Cui
2000; Campana 2000; Lamb \& Miller 2001, 2003; ), rotation, precession and epicyclic frequency
models (Stella \& Vietri 1999; Romanova \& Kulkarni 2009; Bachetti et al. 2010), disk-oscillation and resonance models (Osherovich \& Titarchuk 1999;  Abramowicz \& Kluz\'niak W. 2001, 2003; Kato 2001, 2004;  Urbanec et al. 2010), wave models (Zhang 2004; Li \& Zhang 2005; Rezania \& Samson 2005; Shi \& Li 2009, 2010). The above classification is not strict, because some factors are overlapped in those studies. In most of the models in NS-LMXBs, the characteristic radius is very important, because the changing QPOs frequencies are very dependent on the radius in those models. The location of their production region of the kHz QPOs is discussed and several kinds of places have been suggested, e.g. the surface of the star (such as Romanova \& Kulkarni 2009, Bachetti et al. 2010), the accretion disk (such as Shi \& Li 2009, 2010, Sriram et al. 2011) or the magnetic field lines (such as Zhang 2004, Li \& Zhang 2005).

Zhang (2004) considered the Alfv\'en waves as the source of the kHz QPOs
of the accreting X-ray binaries. Rezania and Samson (2005) also discussed
the MHD turbulence effect coming from the accretion process when the plasma hits the magnetosphere, i.e. the excited resonant shear Alfv\'en waves in a region of enhanced density gradients from collision in the magnetosphere led to the kHz QPOs. Shi \& Li (2009) have considered the two magnetohydrodynamics (MHD) oscillation modes in NS magnetospheres including the effect of gravity and the rotation of a NS as the source of the twin kHz QPOs. A linear relation about the frequencies of the upper QPOs and the lower QPOs was obtained and the model fitted the observation about the change of the upper kHz QPO frequencies and the frequency difference with lower kHz QPOs frequencies well.

In this work, we re-examine the previous MHD model (Shi \& Li 2009) and find
the relation between the kHz QPOs and the accretion rate in NS-LMXBs,
based on the interpretation of the MHD waves of Shi \& Li (2009).
Unlike the result of others, we find that the Alfv\'en-like transverse waves only exist under special conditions. We start in section 2 with the new MHD model and give the solutions of the dispersion equation on the MHD wave frequencies. Then, we compare our results with observations in section 3. Finally, we make discussion in section 4 and summarize our result in section 5.

\section{The MHD model}
In this section, we consider that a steady standard $\alpha$-disk of Shakura \& Sunyaev (1973) in a NS-LMXB is truncated by the stellar magnetosphere,
and the plasma is accreted to the polar cap along the magnetic field (e.g., Ghosh et al. 1977, Elsner \& Lamb 1977). As shown in Fig. 1, in the
accretion process, the plasma hits the magnetic field lines and compresses the primary polar magnetic
field which might lead to a deformation of the magnetic field and some instability (Elsner \& Lamb 1977). The MHD waves produced at the magnetosphere radius from a small perturbation lead to the kHz QPOs.

\subsection{The MHD Waves}

\begin{figure}[h]
\begin{center}
\includegraphics[width=0.5\columnwidth]{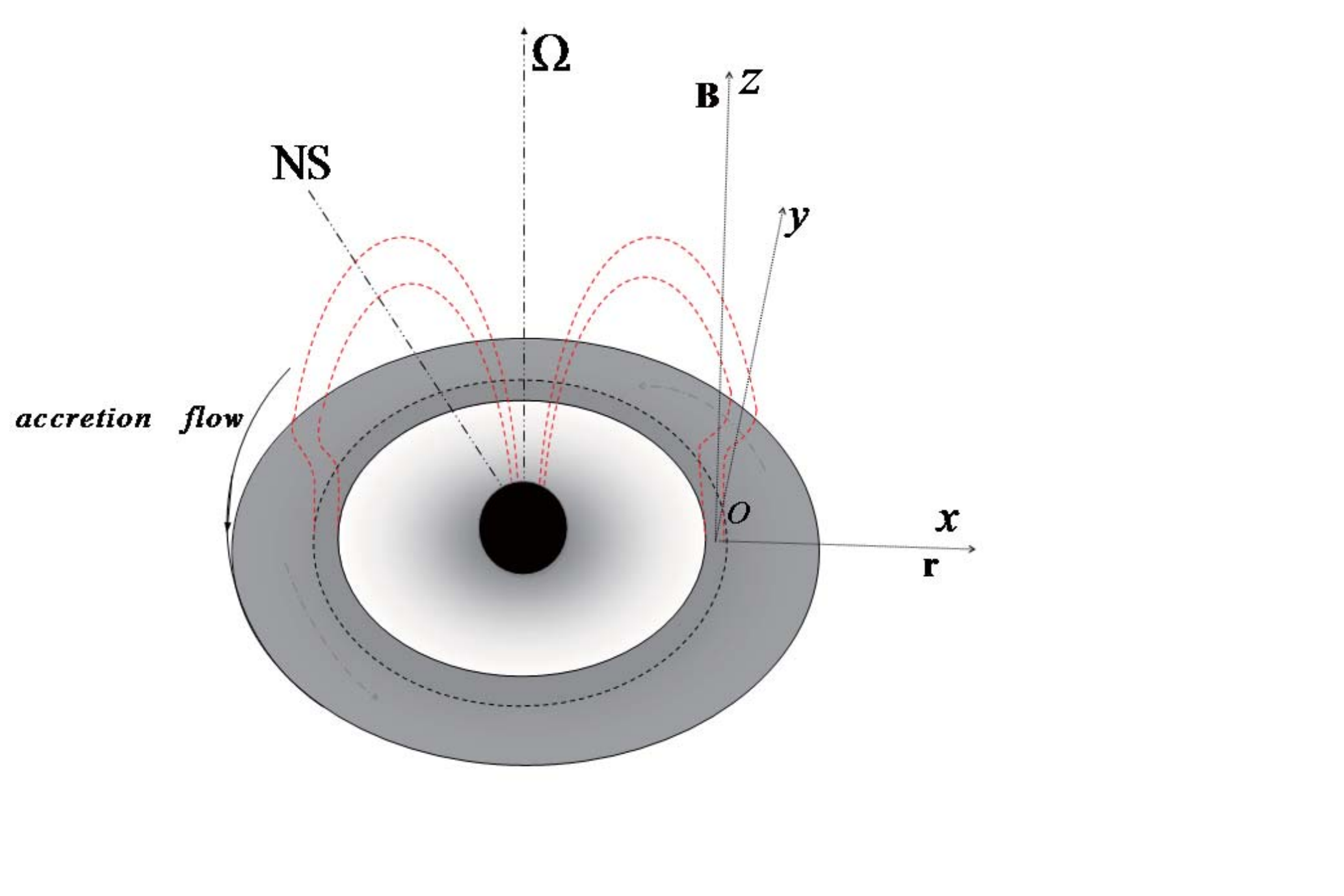}
  \caption{A sketch of the accretion process in NS-LMXBs. The rectangular coordinate system in the
  corotation reference frame is located at the magnetosphere radius and the dotted lines represent the deformation
  of the dipole magnetic field.} \label{fig1}
 \end{center}
\end{figure}

We start from a balance by gravity, barometric pressure, magnetic pressure for the plasma in the border between the magnetosphere and the steady thin accretion disk rotating around the NS; this gives a definition of the magnetosphere radius of the LMXBs. The balance equation can be expressed as follows,
\begin{equation}
\label{eq1}
\rho_0 \frac{\partial \bm u_0 }{\partial t} + \rho _0 \bm u_0 \cdot
\nabla \bm u_0 = - \nabla P_0 + \textstyle{1 \over \mu }(\nabla \times
\bm B_0 )\times \bm B_0 + 2\rho _0 \bm u_0 \times \bm \Omega + \rho _0 \Omega ^2 \bm r_0 -
\rho _0 (\bm \Omega \cdot \bm r_0 )\bm \Omega - \rho _0 \textstyle{{GM} \over {r_0
^3}}\bm r_0,
\end{equation}
where ${\bm u}$ is the plasma velocity, $\mu$ the vacuum magnetic conductivity, $P$ the barometric pressure, $\bm{B}$ the magnetic field, $\bm r$ the displacement from the NS, $\rho$ the plasma density, $G$ the gravitational constant, $M$ the mass
of the NS and ${\bm \Omega}$ the angular velocity of the NS, respectively. The subscript ``0" denotes variables in the equilibrium state and the bold italic expresses vectors. It is very different from Shi \& Li (2009) that now we consider that the magnetic field and the density of the plasma in the magnetosphere are not uniform, but we consider the same initial conditions of Shi \& Li (2009) that the balance is steady at first, i.e.,
$\frac{\partial \rho _0 }{\partial t}=0$, $ \frac{\partial P_0 }{\partial t} = 0$,
$\frac{\partial { {\bm B}}_{\rm {\bf 0}} }{\partial t} = \bm 0$ and $\frac{\partial
{ {\bm \Omega }}}{\partial t} = \bm 0$.
We consider that the plasma is ideal conductor, so the vacuum electroconductivity $\sigma\rightarrow \infty$, and then we can obtain the equation according to the Faraday principle of electromagnet induction as follows,
\begin{equation}
\label{eq2} \frac{\partial { {\bm B}}_{ { 0}} }{\partial t} =
({{\bm B}}_{ { 0}} \cdot \nabla ){ {\bm u}}_{ {0}} - ({ {\bm
u}}_{ { 0}} \cdot \nabla ){{\bm B}}_{ {0}} - (\nabla \cdot
{{\bm u}}_{ {0}} ){{\bm B}}_{ {0}}.
\end{equation}

The continuity and the adiabatic condition are always used in the accretion process and
they can be expressed as follows,
\begin{equation}
\label{eq3} {\partial{\rho_0} \over \partial t}
+\nabla\cdot({\rho_0}{{\bm u_0}})=0,
\end{equation}

\begin{equation}
\label{eq4} P_0\rho_0^{-\gamma}={\rm const},
\end{equation}
where $\gamma$ is the adiabatic index. The plasma at the magnetosphere radius is always disturbed by strong excitation (Rezania \& Samson 2005), so now we discuss the MHD waves that are produced from a small disturbance. The disturbed MHD equations are written as,
\begin{equation}
\label{eq5} \rho \frac{\mbox{d}{{\bm u}}}{\mbox{d}t} + \rho {\bm u} \cdot \nabla {\bm u}= -\nabla P + {
{\bm J}}\times { {\bm B}} + 2\rho {\bm u}\times {\bm\Omega } + \rho
{\bm \Omega }\times ({\bm r}\times {\bm \Omega }) - \rho
\textstyle{{GM} \over {r^3}}{\bm r},
\end{equation}

\begin{equation}
\label{eq6} \frac{\partial{\bm{B}}}{\partial t} = \nabla \times
({\bm u}\times {\bm B}) = ({\bm B} \cdot \nabla ){\bm u} - ({\bm u}
\cdot \nabla ){\bm B} - (\nabla \cdot {\bm u}){\bm B},
\end{equation}

\begin{equation}
\label{eq7} {\partial{\rho} \over \partial t} +\nabla\cdot({\rho
{\bm u}})=0,
\end{equation}

\begin{equation}
\label{eq8} P\rho^{-\gamma}={\rm const},
\end{equation}
where $\bm u =\bm u_0 +\bm u_{\rm s} $, $\bm B = \bm B_0 + \bm B_{\rm s}$, $\bm r =\bm r_0 +\bm r_{\rm s}$, $\rho = \rho _0 + \rho _{\rm s}$, $P = P _0 + P _{\rm s}$ and the subscript ``s" denotes the variation of a physical quantity due to the perturbation, except in the sound velocity ($c_{\rm s}$) below. We consider that a small perturbation lead to the MHD waves, so
$u_{\rm s} \ll \left| {\bm{\Omega} \times \bm{r}} \right|$, $B_{\rm s} \ll B_0$, $r_{\rm s} \ll r_0$, $\rho_{\rm s} \ll \rho_0$, $P_{\rm s} \ll P_0$. Combining Equations (1)$\thicksim$(8), we can obtain the equations about the variation of the perturbed physical quantities in the first-order approximation in the corotation reference frame (so $\bm u_0 = \bm 0$):
\begin{equation}
\label{eq9}
\begin{array}{l}
 \rho _0 \frac{\partial \bm u _{\rm s} }{\partial t} = - \nabla P_{\rm s} +
\textstyle{1 \over \mu }[(\nabla \times \bm B _0 )\times
\bm B _{\rm s} + (\nabla \times \bm B _{\rm s} )\times
\bm B _0 ] + 2\rho _0 \bm u _{\rm s} \times
\bm \Omega + \rho _0 \Omega ^2\bm r _{\rm s} + \\
 \rho _{\rm s} \Omega ^2\bm r _0 - \rho _0 (\bm \Omega \cdot
\bm r _{\rm s} )\bm \Omega - \rho _{\rm s} (\bm
\Omega \cdot \bm r _0 )\bm \Omega
 - \rho _{\rm s} \textstyle{{GM} \over {r_0^3 }}\bm r _0 - \rho _0
GM(\textstyle{{\bm r _{\rm s} } \over {r_0^3 }} -
\textstyle{{3\bm r _0 \cdot \bm r _{\rm s} \bm
r _0 } \over {r_0^5 }}), \\
 \end{array}
\end{equation}

\begin{equation}
\label{eq10}
\frac{\partial \bm B _{\rm s} }{\partial t} = (\bm B _0
\cdot \nabla )\bm u _{\rm s} - (\nabla \cdot \bm u _{\rm s}
)\bm B _0 - (\bm u _{\rm s} \cdot \nabla )\bm
B _0,
\end{equation}

\begin{equation}
\label{eq11}
\frac{\partial \rho _{\rm s} }{\partial t}= - \nabla \rho _0 \cdot \bm
u _{\rm s} - \rho _0 \nabla \cdot \bm u _{\rm s},
\end{equation}

\begin{equation}
\label{eq12}
P_{\rm s}= {\gamma P_0\over {\rho_0}} \rho_{\rm s}.
\end{equation}

Unlike Equation (13) in Shi \& Li (2009), Equation (9) includes the previously neglected term
$-\textstyle{{3\bm r _0 \cdot \bm r _{\rm s} \bm
r _0 } \over {r_0^5 }}$ that comes from $\textstyle{{\bm r _0 + \bm r _{\rm s} } \over {\left|
{\bm r _0 + \bm r _{\rm s} } \right|^3}} -
\textstyle{{\bm r _0 } \over {\left| {\bm r _0 }
\right|^3}} \simeq \textstyle{{\bm r _{\rm s} } \over {r_0^3 }} -
\textstyle{{3\bm r _0 \cdot \bm r _{\rm s} \bm
r _0 } \over {r_0^5 }}$, since sometimes $\textstyle{{3\bm r _0 \cdot \bm r _{\rm s} \bm
r _0 } \over {r_0^5 }}\sim \textstyle{{\bm r _{\rm s} } \over {r_0^3 }}$; in addition we also consider the nonuniform distribution of the density and magnetic field for the plasma in the magnetosphere in those equations.

Differentiating Equation (9) and substituting (12) into it, we obtain
\begin{equation}
\label{eq13}
\begin{array}{l}
 \rho _0 \frac{\partial ^2\bm {u_{\rm s} } }{\partial t^2} \simeq -
\textstyle{\partial \over {\partial t}}(\nabla \frac{\gamma P_0 }{\rho _0
}\rho _{\rm s} ) + \textstyle{1 \over \mu }\frac{\partial }{\partial t}[(\nabla
\times \bm {B_0 } )\times \bm {B_{\rm s} } + (\nabla
\times \bm {B_{\rm s} } )\times \bm {B_0 } ] + 2\rho _0
\textstyle{\partial \over {\partial t}}(\bm {u_{\rm s} } \times
\bm \Omega ) + \\
\rho _0 \Omega ^2\bm {u_{\rm s} } +
\frac{\partial }{\partial t}\rho _s \Omega ^2\bm r _0
 - \rho _0 (\bm \Omega \cdot \bm {u_{\rm s} }
)\bm \Omega - \frac{\partial }{\partial t}\rho _{\rm s}
(\bm \Omega \cdot \bm r _0 )\bm \Omega -
\frac{\partial }{\partial t}\rho _{\rm s} \textstyle{{GM} \over {r_0^3
}}\bm r _0 - \rho _0 \frac{GM}{r_0^3 }(\bm u _{\rm s} -
\textstyle{{3\bm r _0 \cdot \bm u _{\rm s} \bm
r _0 } \over {r_0 ^2}}). \\
 \end{array}
\end{equation}

Now we discuss the physical process in the rectangular coordinate system ($o-xyz$) in the
corotation reference frame (see Fig. 1).  We assume that (i) the accretion disk does not warp, i.e. the compressed magnetic field lines are normal to the disk in the equatorial plane, the magnetic field and the density are only functions of the longitudinal displacement ($r_0$) after being compressed, i.e. $\bm{B_0} = (0, 0, B_0(r_0))$, $\rho_0 = \rho_0(r_0)$, $\bm{\Omega} = (0, 0, \Omega)$; (ii) the MHD waves propagate along the magnetic field lines or they would be dissipated in the disk easily (Shi \& Li 2009), i.e. the wave vector can be expressed as, $\bm k = (0,0,k)$, ($k$ is the wavenumber). The balance equation of the plasma at the magnetosphere radius, Equation (1), can be simplified as,
\begin{equation}
\label{eq14}
0 = - \nabla P_0 + \textstyle{1 \over \mu }(\nabla \times \bm {B_0 } )\times \bm {B_0 } + \rho _0 \Omega ^2\bm
{r_0 } - \rho _0 \textstyle{{GM} \over {r_0 ^3}}\bm {r_0},
\end{equation}
and the further simplified form is,
\begin{equation}
\label{eq15}
\Omega ^2r_0 - \textstyle{{GM} \over {r_0 ^3}}r_0 = (\frac{c_{\rm s}^2}{\rho _0
}\frac{\partial \rho _0 }{\partial r} +
\frac{1}{\mu _0 \rho _0 }B_0 \frac{\partial B_0 }{\partial r})\mid_{r=r_0},
\end{equation}
where ${c_{\rm s} = \sqrt{\frac{\gamma P_0}{\rho_0}}}$ is the sound velocity.

 After substituting Equations (10)$\sim$(12) and (15) into Equation (13), we can obtain,
 \begin{equation}
 \label{eq16}
 \begin{array}{l}
 \rho_0 \frac{\partial^2 \bm u_{\rm s}}{\partial t^2} \simeq \frac{\gamma P_0}{\rho _0 }[\nabla \rho _0 \times (\nabla \times \bm u_{\rm s} ) + (\nabla \rho
_0 \cdot \nabla )\bm u_{\rm s} + (\bm u_{\rm s} \cdot \nabla )(\nabla \rho _0 ) +
\nabla \rho _0 (\nabla \cdot \bm u_{\rm s} ) + \rho _0 \nabla (\nabla \cdot \bm u_{\rm s} )] \\
 + \textstyle{1 \over \mu }\{\nabla \bm B_0 \times [({\rm {\bf e}} \cdot \nabla
)\bm u_{\rm s} ] + \bm B_0 \nabla \times [({\rm {\bf e}} \cdot \nabla )\bm u_{\rm s} ]
- \nabla (\nabla \cdot \bm u_{\rm s} )\times \bm B_0 - (\nabla \cdot \bm u_{\rm s}
)\nabla \times \bm B_0 - \nabla \times [(\bm u_{\rm s} \cdot \nabla )\bm B_0 ]\}\times
\bm B_0 \\
 + \textstyle{1 \over \mu }(\nabla \times \bm B_0 )\times (\bm B_0 \cdot \nabla
)\bm u_{\rm s} - \textstyle{1 \over \mu }(\nabla \times \bm B_0 )\times (\nabla
\cdot \bm u_{\rm s} )\bm B_0 - \textstyle{1 \over \mu }(\nabla \times \bm B_0 )\times
(\bm u_{\rm s} \cdot \nabla )\bm B_0 + \rho _0 \Omega ^2\bm u_{\rm s} - \rho _0 (\Omega \cdot \bm u_{\rm s} )\bm \Omega\\
- \rho _0 \frac{GM}{r_0^3 }(\bm u_{\rm s} - \textstyle{{3r_0 \cdot \bm u_{\rm s}\bm r_0
} \over {r_0 ^2}}) + 2\rho _0 \textstyle{\partial \over {\partial
t}}(\bm u_{\rm s} \times \bm \Omega )
 - (\nabla \rho _0 \cdot \bm u_{\rm s} + \rho _0 \nabla \cdot \bm u_{\rm s})
£\mbox{[}\Omega ^2\bm r_0 - (\bm \Omega \cdot \bm r_0 )\bm \Omega - \textstyle{{GM} \over
{r_0^3 }}\bm r_0 - \nabla (\frac{\gamma }{\rho _0 }P_0 )\mbox{]} ,\\
 \end{array}
 \end{equation}
where $\bf{e}$ is the unit vector that has the same direction with $\bm B_{\rm 0}$. We then carry
out Fourier transformation ($f \rightarrow f e^{i {\bm k \cdot \bm r} - i \omega t}$) for Equation (16) and simplify the result, and obtain the following dispersion equations:
\begin{equation}
\label{eq17}
( - \omega ^2 - \Omega ^2 - 2\omega _{\rm k}^2 \mbox{ + }k^2V_{\rm A}^2 + m)u_{{\rm s}x}
= - ik\textstyle{1 \over {\mu \rho _0 }}B_0 \frac{\partial B_0
}{\partial x}u_{{\rm s}z} + ik(\gamma - 1)\frac{c_{\rm s}^2 }{\rho _0 }\frac{\partial
\rho _0 }{\partial x}u_{{\rm s}z} - i2\omega \Omega u_{{\rm s}y},
\end{equation}

\begin{equation}
\label{eq18}
( - \omega ^2 - \Omega ^2 + \omega _{\rm k}^2 \mbox{ + }k^2V_{\rm A}^2 )u_{{\rm s}y}
= - i2\omega \Omega u_{{\rm s}x},
\end{equation}

\begin{equation}
\label{eq19}
( - \omega ^2 + \omega _{\rm k}^2 \mbox{ + }k^2c_{\rm s}^2 )u_{{\rm s}z}
= (\frac{c_{\rm s}^2}{\rho _0
}\frac{\partial \rho _0 }{\partial x} +
\frac{1}{\mu _0 \rho _0 }B_0 \frac{\partial B_0 }{\partial x})i k u_{{\rm s}x},
\end{equation}

\begin{equation}
\label{eq20}\begin{array}{lll}
m = -[ \frac{\partial {}}{\partial r}(\frac{c_{\rm s}^2}{\rho _0
}\frac{\partial \rho _0 }{\partial r} +
\frac{1}{\mu _0 \rho _0 }B_0 \frac{\partial B_0 }{\partial r})]\mid_{r=r_0},
 \end{array}
\end{equation}
where ${V_{\rm A} = \sqrt{\frac{B_0^2}{{\mu \rho_0}}}}$ is the Alfv\'en velocity and ${\omega_{\rm k} = \sqrt{\frac{GM }{r_0^3}}}$ the Kepler angular frequency. If the magnetic field and the plasma are uniform, $\frac{\partial B_0}{\partial x}=\frac{\partial B_0}{\partial r}=0$ and $\frac {\partial \rho_0}{\partial x}=\frac {\partial \rho_0}{\partial r}=0$ (so $m=0$), this magnetosphere radius is also the corotation radius of the NS from Equation (15), and the frequencies of the MHD Alfv\'en-like waves are,
\begin{equation}
\label{eq21}
\omega ^2 =  k^2V_{\rm A} ^2 + \frac{1}{2}\Omega ^2 \pm \frac{1}{2}\sqrt
{\Omega ^4 + 16k^2V_{\rm A} ^2\Omega ^2}.
\end{equation}

Since the coefficients of the highest order derivatives in Eq. (16) are not small, we cannot make expansions for small coefficients nor assume that these coefficients are equal to 0. Therefore our approximation is a classical method when the total energy is greater than the potential energy in all space, compared with the WKB approximation method.

In the case $m \neq 0$, several solutions are obtained from Equations (17) $\sim$ (20).
\begin{enumerate}
\item If $u_{{\rm s}x}=0, u_{{\rm s}y}=0$ and $u_{{\rm s}z} \neq 0$,
it is a sound-like wave solution, $\omega ^2\mbox{ = }\omega _{\rm k}^2 + k^2c_{\rm s}^2$ on condition that $\gamma c_{\rm s}^2 \frac{\partial \rho _0 }{\partial x} = \textstyle{1 \over \mu
}B_0 \frac{\partial B_0 }{\partial x} + c_{\rm s}^2 \frac{\partial \rho _0
}{\partial x}$.

\item If $u_{{\rm s}x}=0, u_{{\rm s}z}=0$ and $u_{{\rm s}y} \neq 0$, it is a Alfv\'{e}n-like wave solution,
$\omega ^2 = \omega _{\rm k}^2 + k^2 V_{\rm A}^2 - \Omega ^2$, on condition that the Coriolis force is neglected.

\item If $u_{{\rm s}x} \neq 0, u_{{\rm s}y} \neq 0$ and $u_{{\rm s}z} = 0$, i.e. the magnetosphere radius is also the corotation radius, we can obtain the MHD waves as,
$\omega ^2 =  k^2V_{\rm A} ^2 + \frac{1}{2}\Omega ^2 \pm \frac{1}{2}\sqrt
{\Omega ^4 + 16k^2V_{\rm A} ^2\Omega ^2}$, that revert to Equation (21).

\item If $u_{{\rm s}x} \neq 0, u_{{\rm s}y} \neq 0$ and $u_{{\rm s}z} \neq 0$, we can obtain the dispersion equation from Equations (17) $\sim$ (20) as follows,
\end{enumerate}
\begin{equation}
\label{22} \begin{array}{lll}
( - \omega ^2 - \Omega ^2 - 2\omega _{\rm k}^2 \mbox{ + }k^2V_{\rm A}^2 + m)(- \omega ^2 - \Omega ^2 + \omega _{\rm k}^2 \mbox{ + }k^2V_{\rm A}^2)(- \omega ^2 + \omega _{\rm k}^2 \mbox{ + }k^2c_{\rm s}^2)
-4 \Omega ^2 \omega^2 (- \omega ^2 + \omega _{\rm k}^2 \mbox{ + }k^2c_{\rm s}^2)\\
=k^2 r_0^2 (\Omega ^2 - \omega _{\rm k}^2)^2 (- \omega ^2 - \Omega ^2 + \omega _{\rm k}^2 \mbox{ + }k^2V_{\rm A}^2) - (\Omega ^2 - \omega _{\rm k}^2) (- \omega ^2 - \Omega ^2 + \omega _{\rm k}^2 \mbox{ + }k^2V_{\rm A}^2)\gamma k^2 c_{\rm s}^2 \frac{r_0}{\rho_0} (\frac{\partial \rho_0}{\partial r}\mid_{r=r_0})
\end{array}
\end{equation}

The above solutions of the dispersion equations are special solutions under special conditions except the ones of Equation (22), so only the solutions of Equation (22) are ordinary ones and we consider them as the source of the kHz QPOs in NS-LMXBs. Because there is not a concise analytic solution for Equation (22), we will discuss the numerical solutions below.

\subsection{Magnetosphere radius}

The magnetosphere radius is the characteristic radius of the kHz QPOs in this model. Since Lamb et al. (1973) gave a definition about the magnetosphere radius in the globular accretion process onto a compact star, many authors have
explored the outer boundary of the magnetosphere of a pulsar (e.g. Cui 1997) and they obtained not identical but analogous conclusions. The magnetosphere radius obtained by Lamb et al. (1973) is written as,
\begin{equation}
\label{eq23}
r_{\rm m} \simeq 2.29\ast 10^6 \mu_{26}^{4/7} \dot{M}_{16}^{-2/7} m_{1.4\odot}^{-1/7}\ {\rm cm},
\end{equation}
where $r_{\rm m}$ is the magnetosphere radius in the accretion process, $\mu$ the magnetic moment of the star, $\dot{M}$ the accretion rate, $ m$ the mass of the NS   and the subscripts ``26", ``16", ``$1.4\odot$" express the quantities in units of $10^{26}\ {\rm G \cdot cm^3 }$, $10^{16}\ {\rm g/s}$, 1.4 times the mass of the sun, respectively.

McCray \& Lamb (1976) considered that the magnetic pressure could resist the falling spherical plasma layer and they found a magnetosphere radius ranging from $1.3\times10^8\ {\rm cm}$ to $7\times10^8\ {\rm cm}$. Elsner \& Lamb (1977) continued to discuss the spherical accretion process and they considered that the central star rotates sufficiently slowly. In addition a lot of results about the magnetosphere radius close to $10^8\ {\rm cm}$ were calculated if the quantities ($\mu\sim 10^{30}\ {\rm G\cdot cm^3}$, $\dot{M}\sim 10^{16} {\rm g/s} $) were adopted when many conditions, such as the radiation pressure, the different structures of the magnetic field were considered (Davidson \& Ostriker 1973; Burnard et al. 1983; Mitra 1992; Li \& Wang 1995; Wang 1996; Cui 1997; Weng \& Zhang 2011; Shakura et al. 2012).

Long et al. (2005) considered the disk accretion onto the magnetized star with a dipole magnetic field and found by simulation that the magnetosphere radius is half the one from Elsner \& Lamb (1977). Recently Kulkarni \& Romanova (2013) made a three-dimensional MHD simulations of magnetospheric accretion at a quasi-equilibrium state, in which the gravitational, centrifugal
and pressure gradient forces are in balance. Then a different dependence of the magnetosphere radius on the accretion rate of the LMXBs, the magnetic moment, the mass and the radius of the NS is found as,
\begin{equation}
\label{eq24}
r_{\rm m} \approx {2.50}\times 10^6 \mu_{26}^{2/5} \dot{M}_{16}^{-
1/5} m_{1.4\odot}^{-1/10} R_6^{3/10}\ {\rm cm},
\end{equation}
where $R$ denotes the radius of the NS in units of $10^6\ {\rm cm}$. The more gradual changing trend
 with $\mu$ and $\dot{M}$ comes from the compression of the magnetosphere by the disk matter, which leads to the non-dipole magnetic field of the external magnetosphere.

We also consider that a central NS with a dipole magnetic field, whose mass is $m$, accretes plasma from the standard thin disk, in which the density of the plasma is,
 \begin{equation}
\label{eq25}
\rho = 3.1\times 10^{-8}\alpha ^{ - 7 / 10} \dot{M}_{16}^{11/ 20}
m_{\odot}^{ 5 / 8} R_{10}^{- 15 / 8} f^{11 / 5}\ {\rm g\ cm^{-3}},
\end{equation}
where ${f = (\mbox{1} - \sqrt {\frac{R}{r}} )^{1 / 4}}$, $\alpha$ is the viscosity parameter (Frank et al. 2002).

We can obtain the barometric pressure ${P = \frac{\rho K T}{m_{\rm p}}}$ and $T$ is temperature in the standard thin disk, where $K$ is Boltzmann constant and $m_{\rm p}$ is mass of the proton. We consider the balance of the plasma by magnetic pressure, barometric pressure and collision, and the magnetosphere radius can be estimated according to, ${\frac{B^2}{8 \pi} \simeq P + \rho u^2}$ (Eslner \& lamb 1977; Romanova et al. 2002).
Due to $P \gg\rho u_r^2$ in the standard thin disk when the radius ranges from $10^6\ {\rm cm}$ to $10^{10}\ {\rm cm}$, the balance condition can be simplified as, ${P \simeq \frac{B^2}{8 \pi}}$. The magnetosphere radius can be estimated when the dipole magnetic field is adopted,
\begin{equation}
\label{eq26}
r_{\rm m} \simeq 6.64\times 10^5 \alpha^{4 / 15} \mu_{26}^{16 / 27}
\dot{M} _{16}^{ - 34 / 135} m_{1.4\odot}^{ - 7 / 27} f^{ - 136 / 135}\ {\rm cm}.
\end{equation}
We have also made the strict computation on the magnetosphere radius by the equation, ${|\nabla (\frac{B^2}{8 \pi}) | \simeq |\nabla P |}$, and the result is very close to the above magnetosphere radius.

If we select the characteristic value, i.e., $m = 1.4 M_\odot$, $R = 10^6\ {\rm cm}$, and $\alpha = 0.1$ (King et al. 2007), all the magnetosphere radius can be simplified. The magnetosphere radius in the spherical accretion process can be simplified from Equation (23) as,
\begin{equation}
\label{eq27}
r_{\rm ms} \simeq 2.29 \times 10^6 \dot{M} _{16}^{-2/7}  B_{\ast8}^{4 / 7}{\rm cm},
\end{equation}
where $B_{\ast 8}$ denotes the magnetic flux density at the surface of the NS in the unit of $10^8\ \rm G$.
After the parameters being substituted into Equation (26) the magnetosphere radius is simplified as,
\begin{equation}
\label{eq28}
r_{\rm mA} \simeq 3.60 \times 10^5 \dot{M}_{16}^{-34/135} B_{\ast8}^{16/27} f^{ - 136 / 135} {\rm cm},
\end{equation}
 which we call ``model A" when the radius $r_{\rm mA}$ is used. We also simplify Equation (24) as,
\begin{equation}
\label{eq29}
r_{\rm mB} \simeq {2.50} \times 10^6 \dot{M}_{16}^{-1/5} B_{\ast8}^{2/5}\ {\rm cm},
 \end{equation}
which we call ``model B" when the compressed magnetosphere radius ($r_{\rm mB}$) is considered.

As shown in Figure 2, the magnetosphere radius in the spherical accretion process ($r_{\rm ms}$) is much larger and covers a larger range than the others.
The magnetosphere radius in model A approaches the radius of a NS when $\dot{M}$ is high enough or the magnetic field at the surface of NS is weak enough. The compressed magnetosphere radius ($r_{\rm mB}$) changes slower than the other two radii ($r_{\rm ms}$ and $r_{\rm mA}$), which reveals the clear differences between the compressed magnetic field and the dipolar magnetic field. In this study we only discuss the accretion process of the plasma in accretion disk of LMXBs, so the magnetosphere radius of Equation (27) for spherical accretion will not be used hereafter. The innermost stable circular orbit (ISCO) is the smallest stable circular orbit of accretion plasma and the plasma will fall to the central NS after it passes through the orbit due to a dynamical instability for circular geodesics in general relativity. The ISCO of the accretion plasma around a NS with the mass ($1.4M_\odot$) is marked as the dash lines in Figure 2 according to its radius (${\frac{6Gm}{c^2}}$). Inside ISCO, the assumption of the standard disk model breaks down, so we assume that the disk is truncated at ISCO at high $\dot{M}$.

\begin{center}
\begin{figure*}[h]
\label{fig2}
  \includegraphics[width=1\columnwidth]{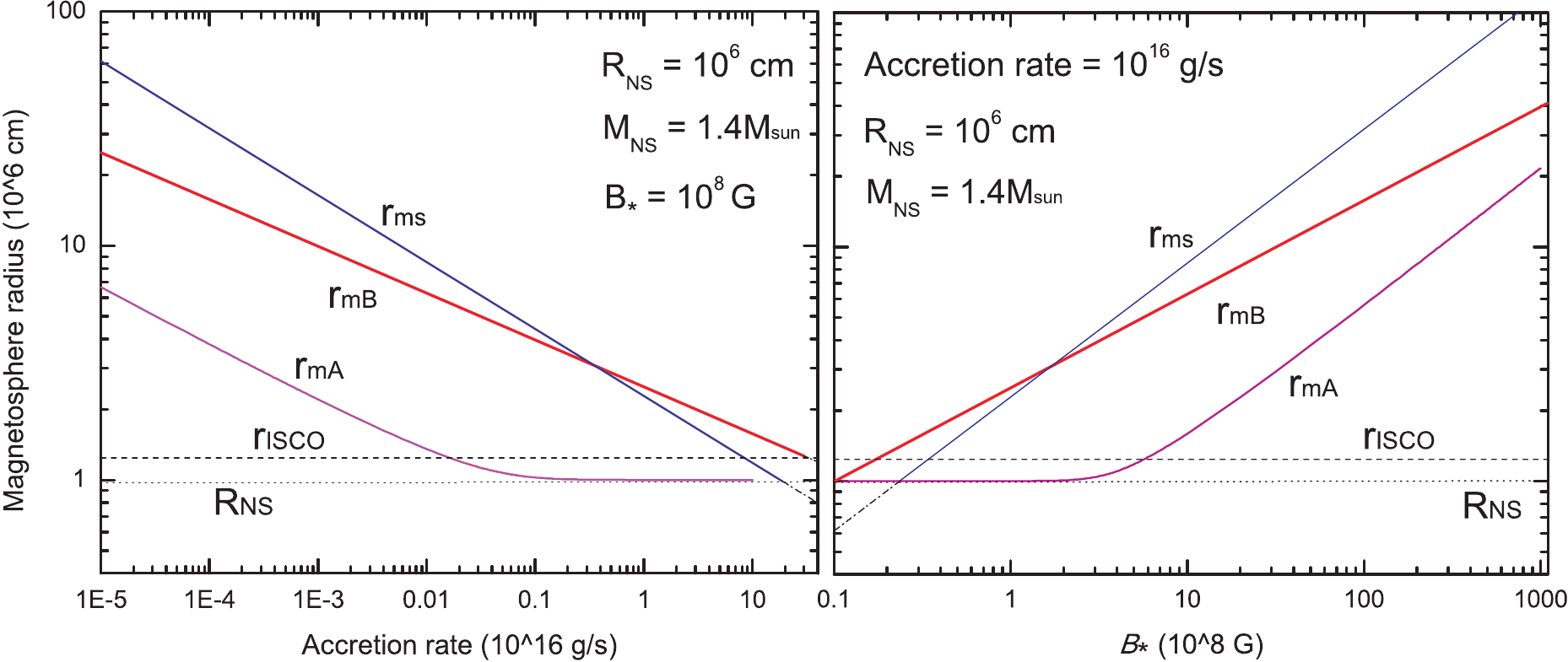}
  \caption{ The relation between the magnetosphere radius and the accretion rate of LMXBs or the surface magnetic field of NS from three models.}
\end{figure*}
\end{center}

\subsection{The general solutions in two kinds of magnetic configurations}
We suppose that the MHD waves at the magnetosphere radius are the origin of the kHz QPOs, so numerical solutions of Equation (22) should be obtained according to the different parameters, $B_*, \dot{M}, k$, and for the characteristic value of NS-LMXBs, i.e., $m = 1.4 M_\odot$, $R = 10^6\ {\rm cm}$, and $\alpha = 0.1$ (King et al. 2007). In the two models (i.e. model A and model B), two types of magnetic field configurations are adopted, distinguished by that the magnetic field is compressed or not.

In model A, the standard $\alpha$-disk contacts with the dipolar magnetic field at the magnetosphere radius and an equilibrium state is reached. We substitute the balance equation, $\frac{c_{\rm s}^2}{\rho _0
}\frac{\partial \rho _0 }{\partial r} = \Omega ^2r_0 - \textstyle{{GM} \over {r_0 ^3}}r_0- \frac{1}{\mu _0 \rho _0 }B_0 \frac{\partial B_0 }{\partial r}$ (i.e. Equation (15)), into Equation (22) and obtain the equation when the dipolar magnetic field is considered,
\begin{equation}
\label{30} \begin{array}{lll}
( - \omega ^2 - \Omega ^2 - 2\omega _{\rm k}^2 \mbox{ + }k^2V_{\rm A}^2 + m )(- \omega ^2 - \Omega ^2 + \omega _{\rm k}^2 \mbox{ + }k^2V_{\rm A}^2)(- \omega ^2 + \omega _{\rm k}^2 \mbox{ + }k^2c_{\rm s}^2)
-4 \Omega ^2 \omega^2 (- \omega ^2 + \omega _{\rm k}^2 \mbox{ + }k^2c_{\rm s}^2)\\
=(1-\gamma)k^2 r_0^2 (\Omega ^2 - \omega _{\rm k}^2)^2 (- \omega ^2 - \Omega ^2 + \omega _{\rm k}^2 \mbox{ + }k^2V_{\rm A}^2) -3 \gamma k^2V_{\rm A}^2(\Omega ^2 - \omega _{\rm k}^2) (- \omega ^2 - \Omega ^2 + \omega _{\rm k}^2 \mbox{ + }k^2V_{\rm A}^2).
\end{array}
\end{equation}
In Equation (30) the Kepler angular frequency, Alfv\'en velocity, sound velocity can be obtained as follows, $\omega_{\rm k} = 13628.4 r_{\rm mA,6}^{1.5}\ {\rm Hz}$, $V_{\rm A} = 2.56\times10^7 B_8 \dot{M}_{16}^{-11/40} r_{\rm mA,6}^{-\frac{33}{16}} f_{\rm A}^{-11/10} \ {\rm cm/s}$, $c_{\rm s} = 1.29\times10^8 \dot{M}_{16}^{3/20} r_{\rm mA,6}^{-3/8} f_{\rm A}^{11/5}\ {\rm cm/s}$, where $f_{\rm A} = (1-\sqrt{1/r_{\rm mA,6}})^{1/4}$.

In model B, the standard $\alpha$-disk compresses the magnetic field untill an equilibrium state is reached and the relevant physical quantities for Equation (22) are obtained as, $\omega_{\rm k} = {3447.74} \dot{M}_{16}^{3/10} B_{\ast8}^{-3/5}$, $c_{\rm s} = {4.10} \times 10^7 B_{\ast8}^{-3/20} \dot{M}_{16}^{9/40}f_{\rm B}^{3/5}$, where $f_{\rm B} = (1-{{0.63}{ B_{\ast8}^{-1/5} \dot{M}_{16}^{1/10} }})^{ 1 / 4}$. Because the dipole magnetic field is compressed within the magnetosphere ($\sigma \rightarrow \infty$), we adopt an approximation about the magnetic field structure that all the magnetic field lines, which are compressed between the magnetosphere radius ($r_{\rm mB}$) and the magnetosphere radius of the uncompressed magnetic field ($r_{\rm mA}$), are uniform. During the compressing process the magnetic flux remains unchanged, so the magnetic field can be estimated from the equation,
\begin{equation}
\label{eq31}
\int_{r_{\rm mA} }^{r_{\rm mB} } {B_0^{'} 2\pi rdr} = \int_{r_{\rm mA} }^\infty {B_{\rm p} 2\pi rdr},
\end{equation}
 where $B_0^{'}$ is the magnetic flux density of the compressed magnetic field and $B_{\rm p}$ is the one of the uncompressed dipole magnetic field in the equatorial plane. The Alfv\'en velocity can be obtained as,
\begin{equation}
\label{eq32}
V_{\rm A}  = {1.51 \ast 10^8} B_{\ast8}^{11/8} \dot{M}_{16}^{-37/80} (r_{\rm mB,6}^2-r_{\rm mA,6}^2)^{ - 1}r_{\rm mA,6}^{-1}f_{\rm B}^{-11/10}\ {\rm cm / s}.
\end{equation}

In order to derive the general solutions we should derive the secondary derivative of the density and the magnetic field (i.e. $m$ in Equation (20)) in model A and model B, we assume that the condition on the balance of plasma (i.e. Equation (15)) is extended to the vicinity of the magnetosphere radius, so ${m\simeq-{\partial (\Omega ^2r- \textstyle{{GM} \over {r^3}}r) }/{\partial r} =-\Omega ^2- 2\omega_{\rm k}^2}$.
Besides that, we use the distribution of the density of the $\alpha$-disk (${\frac{\partial \rho}{\partial r}}$) in Equation (22) in model B or the distribution of the dipolar magnetic field (${\frac{\partial B}{\partial r} }$) in Equation (30) in model A. Finally we obtain the general numerical solutions after the two kinds of magnetosphere radii in Equations (28) and (29) are substituted into Equations (22) and (30), respectively (see Figure 3).

\begin{center}
\begin{figure*}[h]
\label{fig3}
 \includegraphics[width=1\columnwidth]{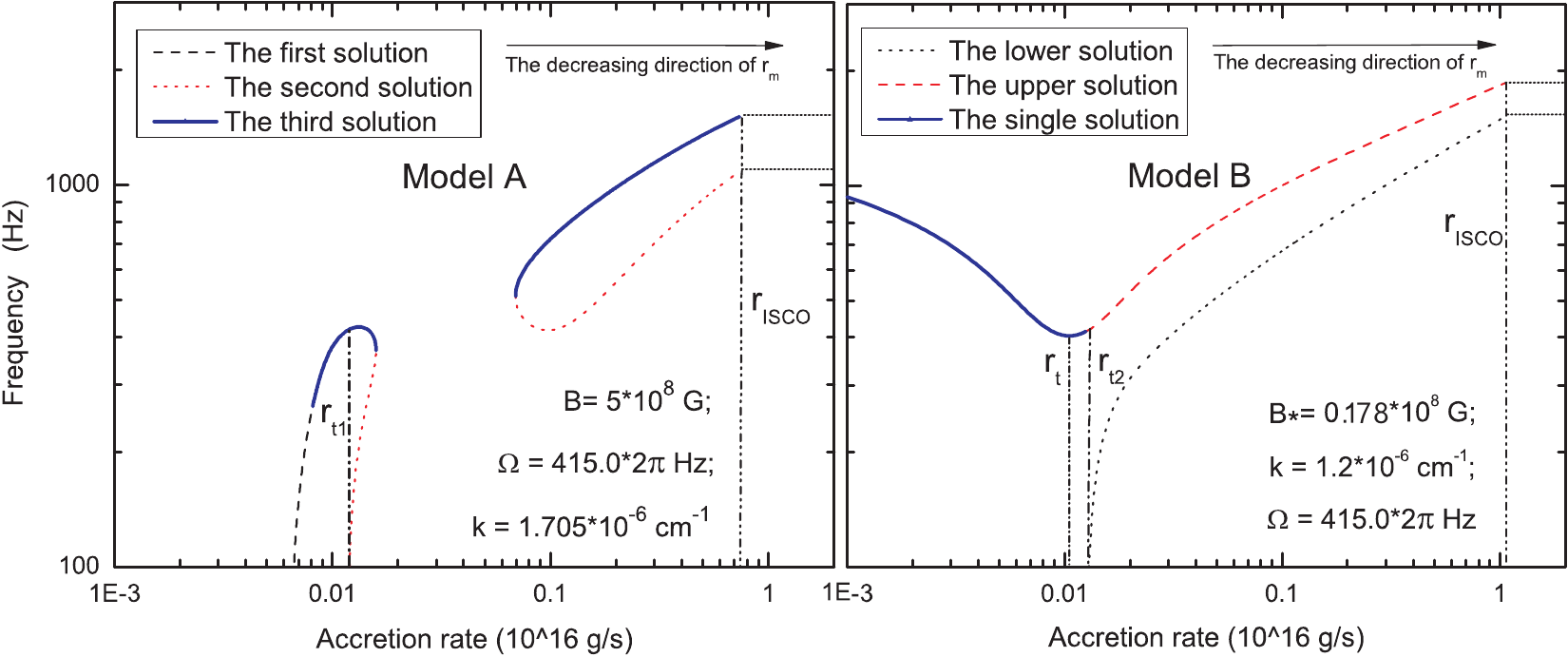}
 \caption{ {\it Left}: the general real solutions of Equation (30) when the dipolar magnetic field is considered in model A; {\it right}: the general real solutions of Equation (22) when the compressed magnetic field is considered in model B.}
\end{figure*}
\end{center}

As shown in Figure 3, we can find that there are twin solutions (the upper solution $\nu_{\rm u}$ and the lower solution $\nu_{\rm l}$) and single solutions both in the two models; the solutions within $r_{\rm t1}$ can be considered as twin solutions and the solutions outside $r_{\rm t1}$ are single solutions in the left panel of Figure 3. In the left panel, the four characteristics of the solutions of Equation (30) in model A are listed as follows. (i) The solutions are divided into two parts, i.e. the left solutions corresponding to low $\dot{M}$ and the right ones corresponding to high $\dot{M}$.
(ii) A transition accretion rate corresponding to the transition radius ($r_{\rm t1}$) can be found by the vertical dash-dotted line; the transition radius is the border between the twin solutions and the single solution for an accretion rate and it is near the corotation radius.
(iii) In the right part there are only twin solutions for an accretion rate.
(iv) $\nu_{\rm u}$ and $\nu_{\rm l}$ increase with $\dot{M}$ until they reach their ceiling frequencies at $\nu_{\rm u}=1508$~Hz and $\nu_{\rm l}=1083$~Hz, due to the restriction of ISCO.

In the right panel of Figure 3, the general real solutions of Equation (22) are obtained and their five characteristics are listed as follows.
(i) The single solution described by the bold solid line is also the upper solution and supposed as the origin of the single kHz QPOs.
(ii) $\nu_{\rm u}$ and $\nu_{\rm l}$ increase with $\dot{M}$ until they reach their ceiling frequencies at $\nu_{\rm u}=1862$~Hz and $\nu_{\rm l}=1521$~Hz when the accretion disk is truncated by ISCO.
(iii) There is a transition point (corresponding to a transition radius $r_{\rm t2}$) that splits the solutions into the twin solutions and the single one; the lower real solutions will disappear when $\dot{M}$ is lower than the transition accretion rate.
(iv) The turning point of the single solutions (corresponding to a transition radius $r_{\rm t}$) separates the decreasing trend from the increasing trend of $\nu_{\rm u}$ and means the changing of the key factor dominating the balance of the plasma.
(v) All the two transition radii ($r_{\rm t}$ and $r_{\rm t2}$) are close to the corotation radius at which the Kepler rotation frequency of the plasma equals to the spin of a NS.

\begin{centering}
\begin{figure*}[htb!]
\label{fig4}
     \includegraphics[width=1\columnwidth]{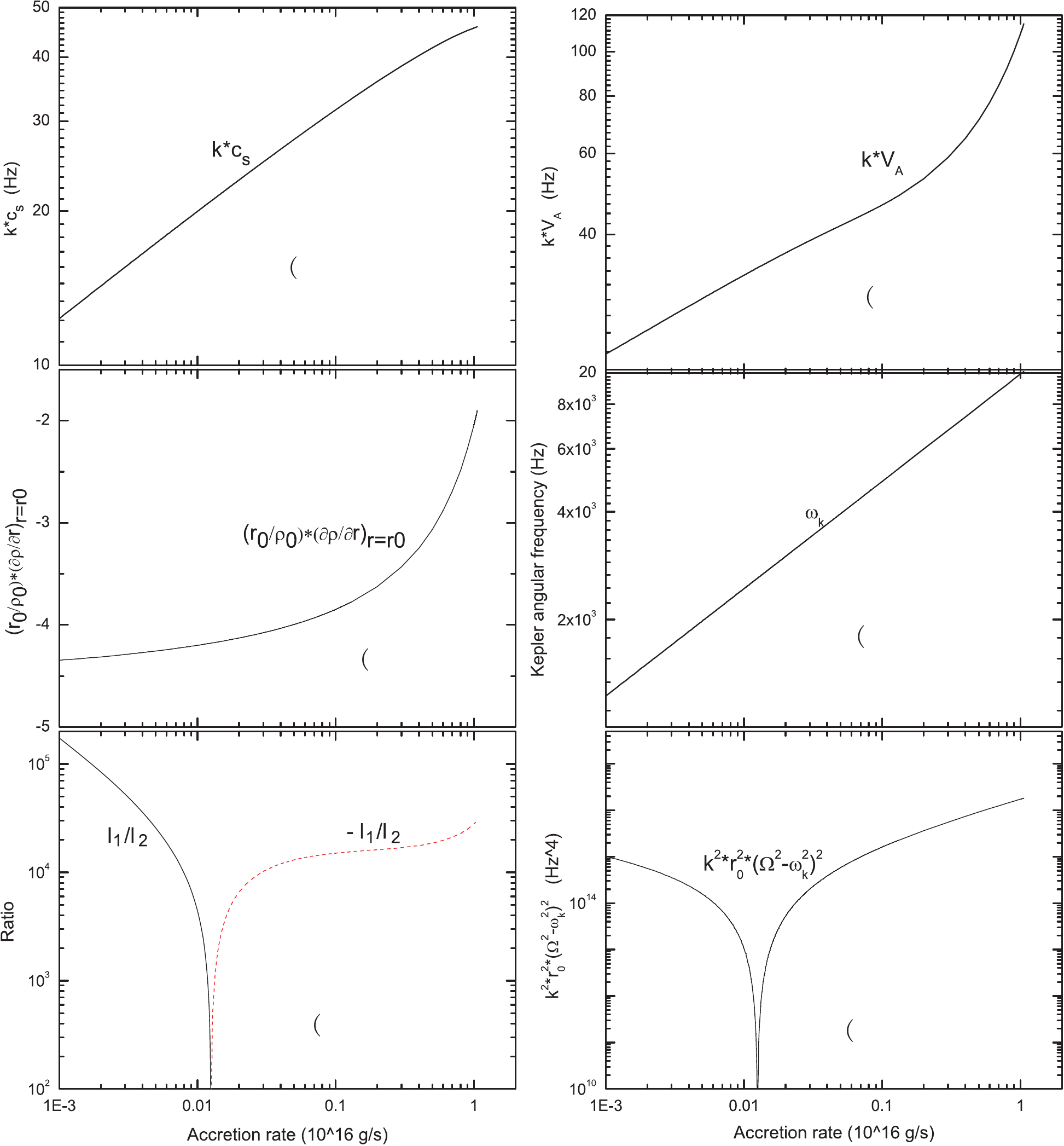}
 \caption{ The relation between the quantities ($k*V_{\rm A},\ k*c_{\rm s}$, $\omega_{\rm k}$, $\frac{r_0}{\rho_0}\frac{\partial \rho}{\partial r}\mid_{r=r_0}$, $k^2r_0^2*(\Omega^2-\omega_{\rm k}^2)^2$ and a ratio) and $\dot{M}$ in model B when the same parameters with the right panel of Figure 3 are adopted.}
\end{figure*}
\end{centering}

The transition accretion rates in the two models (corresponding to $r_{\rm t1}$ and $r_{\rm t2}$) all separate the twin solutions from the single solutions and they originate from the same reason: with the decrease of $\dot{M}$ the quantity ($k^2 v_{\rm A}^2+\omega_{\rm k}^2-\Omega^2$) decreases from a positive value into a negative value and meanwhile the twin solutions change into a single solution, i.e. the lower real solutions disappear. According to the transition condition, $k^2 v_{\rm A}^2+\omega_{\rm k}^2=\Omega^2$, and the balance Equation (15), we can obtain a positive expression ${r_0 k^2 v_{\rm A}^2=(\frac{c_{\rm s}^2}{\rho _0
}\frac{\partial \rho _0 }{\partial r} +
\frac{1}{\mu _0 \rho _0 }B_0 \frac{\partial B_0 }{\partial r})\mid_{r=r_0}}$. In model A, ${\frac{\partial B_0}{\partial r} < 0}$ of the dipolar magnetic field leads to ${\frac{\partial \rho_0}{\partial r} > 0\mid_{r=r_0}}$, i.e., a compressed dense plasma layer. Conversely in model B, ${\frac{\partial \rho_0}{\partial r} < 0}$ of the standard disk density leads to ${\frac{\partial B_0}{\partial r} > 0\mid_{r=r_0}}$, i.e., a compressed stronger magnetic field topology.

The changing trend of the single solutions in model B is different from the one in model A; the main factor can be found from Figure 4 in model B and Equation (22), as follows.
(i) The Kepler angular frequency is very high (${\omega_{\rm k}\simeq50\sim110kv_{\rm A}}$ \& $\omega_{\rm k}\simeq100\sim210 kc_{\rm s}$) and it is the key factor to determine the solutions of Equation (22).
(ii) With the increase of $\dot{M}$, $r_{\rm m}$ decreases and the other variables ($\omega_{\rm k},\ \ k*V_{\rm A},\ k*c_{\rm s}$, $\frac{r_0}{\rho_0}\frac{\partial \rho}{\partial r}\mid_{r=r_0}$) in Equation (22) increase.
(iii) The expression $l_2 \equiv -(\Omega^2-\omega_{\rm k}^2)\gamma k^2c_s^2 \frac{r_0}{\rho_0} \frac{\partial \rho_0}{\partial r}\mid_{r=r_0}$ is smaller than $l_1 \equiv k^2*r_0^2*(\Omega^2-\omega_{\rm k}^2)^2$ except a small section near the turning point (see panel E of Figure 4),
  so Equation (22) can be simplified as,
\begin{equation}
\label{33} \begin{array}{lll}
( \omega ^2 +2\Omega ^2 + 4\omega _{\rm k}^2 \mbox)(- \omega ^2 - \Omega ^2 + \omega _{\rm k}^2 )(- \omega ^2 + \omega _{\rm k}^2)
+4 \Omega ^2 \omega^2 (- \omega ^2 + \omega _{\rm k}^2 )
\simeq k^2 r_0^2 (\Omega ^2 - \omega _{\rm k}^2)^2 (\omega ^2 + \Omega ^2 - \omega _{\rm k}^2).
\end{array}
\end{equation}
(iv) In Equation (33), it can be inferred that the expression $l_1$ dominates the changing trend because all the other expressions keep the changing trend with increasing $\dot{M}$.
 The changing trend of $l_1$ is shown in panel F of Figure 4 and its transition accretion rate is also close to that of the single solution.

We can draw an conclusion that the frequencies of the solutions of Equation (22) will change with the increase of the accretion rate in the same trend with $l_1$.

The negative factor ($1-\gamma$) in Equation (30) in model A from the balance Equation (15) leads to a different changing trend of the frequencies of the single solutions with the accretion rate. The configuration of the magnetic field is related to the distribution of the density by the balance equation. Finally we can conclude that the differences of the configuration of the magnetic field and the distribution of the density lead to the different changing trends of the frequencies of the solutions in model A and B.

\section{Comparison with observations}

\subsection{Twin kHz QPOs}
We take several steps to select the suitable parameters, such as the wavenumber, the surface magnetic field of NSs to match the observations.
\begin{enumerate}
\item Choose the initial value of the magnetic flux density (e.g. $B_{*}=10^8\ {\rm G}$) and the accretion rate (e.g. $\dot{M}=10^{16}\ {\rm {g/s}}$), and then substitute them into $r_{\rm mB}$ or $r_{\rm mA}$, and the detailed expressions on $\omega_{\rm k}, c_{\rm s}, V_{\rm A}$, $m$.
\item The solutions of Equation (22) or (30) can be obtained when different wavenumbers, $k$, are considered.
 \item Select the value of $k$ for which the model predicted frequency difference ($\bigtriangleup \nu=\nu_{\rm u}-\nu_{\rm l}$) is closest to the discovered average value of $\bigtriangleup \nu=\nu_{\rm upper}-\nu_{\rm lower}$ (such as 322 Hz for 4U 0614+09).
 \item The solutions of Equation (22) or (30) can be obtained again with the above $k$, the same accretion rate (e.g. $\dot{M}=10^{16}\ {\rm {g/s}}$) and different values of $B_{*}$.
 \item Select the value of $B_*$, for which the twin solutions are closest to the observed frequencies of the twin kHz QPOs.
\item After substituting the above $k$, $B_*$ and different values of $\dot{M}$ into Equation (22) or (30), we can obtain new solutions.
\item Change the value of $B_*$ slightly and and repeat step 6 until the solutions match the frequencies of most observed twin kHz QPOs.
\item  Change the value of $k$ slightly and repeat step 6 until the solutions match the frequencies of most observed twin kHz QPOs.
\item Repeat the above steps (7) and (8) in turn until the numerical solutions match the maximum number of the observed twin kHz QPOs.
 \item With the last $k$ and $B_*$, we can obtain both $\nu_{\rm u}$ and $\nu_{\rm l}$ changing with $\dot{M}$ in the two models.
 \item With all the parameters determined above (and listed in Table 1), We compute $\dot{M}$ by requiring $\nu_{\rm lower}=\nu_{\rm l}$.
 \item For each $\dot{M}$, we numerically calculate $\nu_{\rm u}$.
 \item Finally in Figure 5, we plot the numerically calculated and observed twin kHz QPOs as functions of $\dot{M}$.
\end{enumerate}

Because the analytic solutions are too complex and their expressions are too long to be used, we just choose the ``best" parameters that match the most data as much as possible by eye-balling.
As shown in Table 1, most of the wavenumbers ($k$) are close to $\rm{1*10^{-6}}\ cm^{-1}$ in Table 1. Rezania and Samson (2005) discussed that the wavelength of the MHD wave in LMXBs was in the order of magnitude of the radius of NS.

\begin{figure*}[!htbp]
\begin{center}
\label{fig5}
\includegraphics[width=0.65\columnwidth]{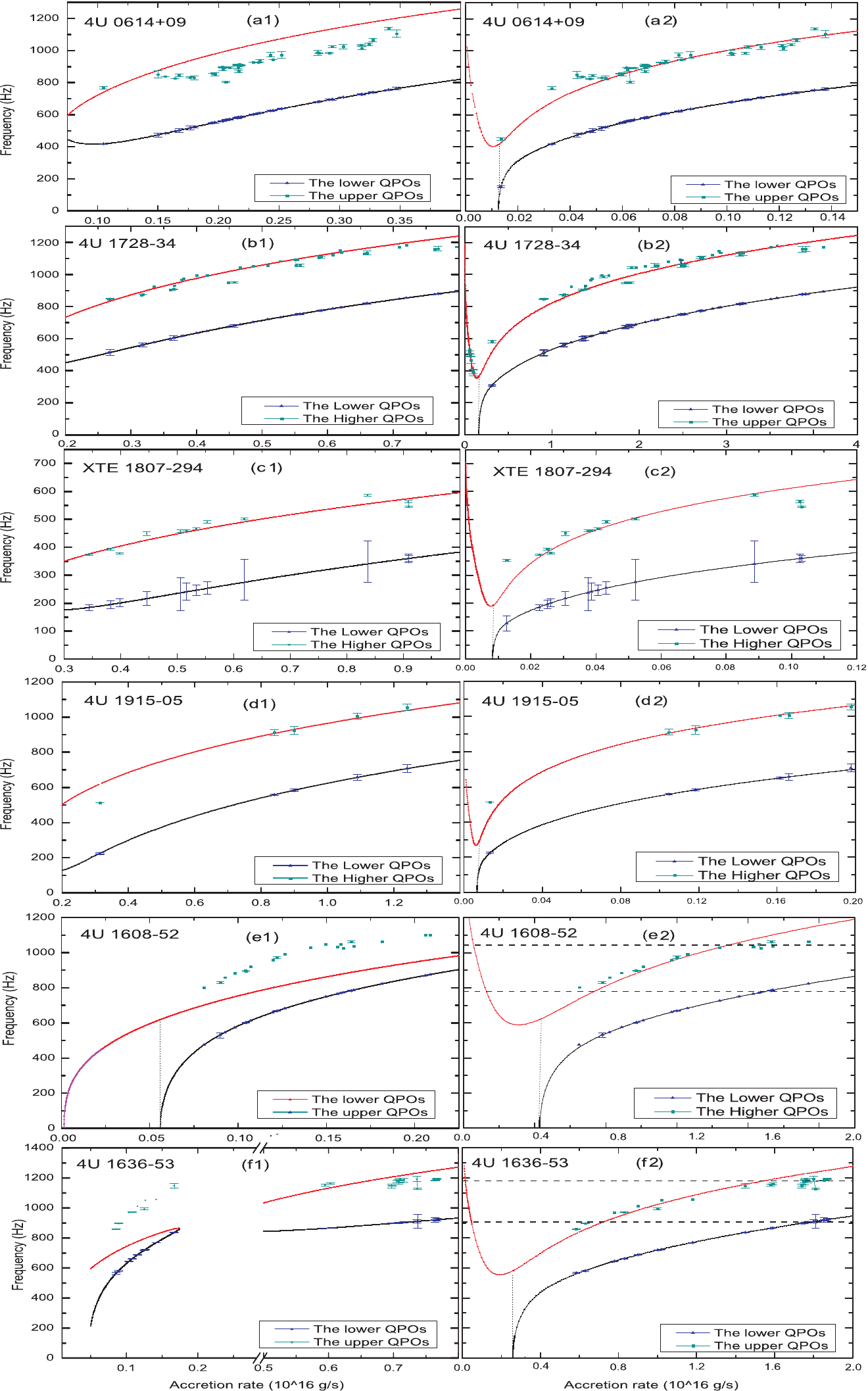}
\caption{ The relations between the frequencies of the twin kHz QPOs, $\nu_{\rm upper}$, $\nu_{\rm lower}$, and accretion rate $\dot{M}$ for
the six sources (for the measured data of 4U 0614+09: van Straaten
et al. 2000; van Straaten et al. 2002; 4U 1636--53: Altamirano et al. 2008; Di
Salvo, M\'endez et \& van der Klis 2003; Jonker, M\'endez \& van der
Klis 2002; Wijands et al. 1997; 4U 1608--52: van Straaten, van
der Klis \& M\'endez 2003; 4U 1915--05: Boirin et al. 2000; 4U 1728--34: van Straaten et al. 2002, Migliari, van der Klis \& Fender 2003; Di Salvo et al. 2001; Jonker, M\'endez \& van der
Klis 2000;  Strohmayer et al. 1996; XTE 1807--294: Linares et
al. 2005; Zhang et al. 2006). {\it left}: The magnetosphere radius from Equation (28) is used and the dipolar magnetic field is considered in model A. {\it right}: The compressed magnetic field is considered and we use the magnetosphere radius (Equation (29)) which is obtained from the simulation of Kulkarni \& Romanova (2013) in model B.}
\end{center}
\end{figure*}

\begin{table*}[h]
 \centering
  \caption{This table lists the parameters used in our numerical computation for two different kinds of magnetosphere radii: the spin $\nu = \frac{\Omega}{2\pi}$, the selected parameters ($k_{\rm A1}, k_{\rm A2}, B_{\ast \rm A}$) for model A, the selected parameters ($k_{\rm B}, B_{\ast {\rm B}}$) for model B, the corotation radius ($r_{\rm co}$) and the transition radius $r_{\rm t}$ for model B in the
corresponding sources respectively. }
\label{t:1}
 \scriptsize{
  \begin{tabular}{@{}lcccccccc@{}}
   \hline
  \hline
  \multirow{2}*{sources} & $\nu$ & $k_{\rm A1}$  & $k_{\rm A2}$ & $ B_{\ast \rm A}$ & $k_{\rm B}$ & $B_{\ast {\rm B}}$ & $r_{\rm co}$ & $r_{\rm t2}$\\
   & $({\rm Hz})$ & $ (10^{-6}\ {\rm cm^{-1}})$ & $ (10^{-6}\ {\rm cm^{-1}})$ & $ (10^8\ {\rm G})$ & $ (10^{-6}\ {\rm cm^{-1}})$ & $ (10^8\ {\rm G})$ &$ (10^6\ {\rm cm})$ & $ (10^6\ {\rm cm})$ \\
 \hline
4U 0614+09 & 415 &  1.71 & $\thicksim$ & 5 &  1.20 & 0.18  & 3.00  & 3.01\\

4U 1728--34 & 363 & 1.52 & $\thicksim$ & 7 &  1.15 & 0.80 & 3.28  & 3.29\\

XTE 1807--294 & 190.6 & 1.03 & $\thicksim$ & 20 &  1.13 & 0.53 & 5.04  & 5.08\\

4U 1915--05 & 270 & 1.35 & $\thicksim$ & 11 & 1.33 &  0.27  & 4.00  & 4.01\\
\hline
4U 1608--52 & 619 & 0.05 & $\thicksim$ & 6 &  1.10 & 0.53  & 2.30  & 2.31\\

4U 1636--53 & 581 & 0.36 & 1.97 & 6 &  1.10 & 0.46  & 2.40  & 2.41\\
\hline
\end{tabular}}
\end{table*}
The data points describe the observational data and the solid lines come from our numerical solutions in Figure 5. In the left panels of Figure 5, we consider the dipole magnetic field and use the magnetosphere radius in model A from Equation (28). In ${\rm a1,\ b1,\ c1}$, suitable values of $\dot{M}$ for several groups of observed QPOs are not found with model A and thus those QPOs are not plotted. The right panels describe the result while the compressed magnetic field is considered and the magnetosphere radius from Equation (29) is used.

In the left panels of Figure 5 we could not find a group of parameters for 4U 0614+09, 4U 1608--52, 4U 1636--53 in order to match the observations and we can only adopt not one but two different $k$ ($k_{\rm a1}, k_{\rm a2}$) for 4U 1636--53 for the best result to match the observations including all the twin kHz QPOs.  The relation between $\nu_{\rm upper}$ and $\dot{M}$ is reproduced well in the right panels of Figure 5 and the result from model B is much better than that from model A.

In the right panels of Figure 5, our numerical solutions in 4U 1608--52 and 4U 1636--53 deviate the observation slightly in the tails of the curves maybe due to ISCO. Barret et al. (2005, 2006) discovered the ceiling of the lower QPO frequency in 4U 1636--53 and 4U 1608--52 in a frequency-count rate diagram.
As shown in Figure 5, however the ceiling of $\nu_{\rm upper}$ in 4U 1636--53 and 4U 1608--52 seem to be clearer than $\nu_{\rm lower}$. This is is different from Barret et al. (2005) and the main reason is that in plotting our results in the figure, $\nu_{\rm l}=\nu_{\rm lower}$ is required to determine $\dot{M}$ and so deviations can only exist in $\nu_{\rm upper}$. Then the magnetosphere radius can be determined with Equation (29).

We can estimate the masses of the two NSs in 4U 1636--53 and 4U 1608--52 if the magnetosphere radius corresponding to the ceiling frequency is ISCO (see Figure 3), i.e., $M_{\rm NS} =\frac{r_2}{r_{\rm 1}}*{1.4M_\odot}$, where $r_1$ is the ISCO of a NS with $1.4M_\odot$ and $r_2=r_{\rm_m}$ in Equation (29) when the QPO frequency reach the ceiling frequency. Then the ceiling frequencies and their errors can be found as follows (see Figure 6):
\begin{enumerate}
\item The histogram of the lower (or upper) QPO frequencies is plotted with the bin size of 20 Hz (slightly larger than the error (about 17 Hz) of the observational data);
\item We look for a peak in the histogram near the high frequency end;
\item We pick all the QPO frequencies if their absolute differences between the peak frequency are less than 40 Hz;
\item The average of those QPO frequencies is considered as the ceiling frequency and their root-mean-square (RMS) is the error of the ceiling frequency.
\end{enumerate}
With the ceiling frequency of either $\nu_{\rm upper}$ or $\nu_{\rm lower}$, we can obtain $\dot{M}$ from the numerical solution for model B. Then with $B_*$ (see Table 1) determined from the fitting (see section 3.1 for details), $r_2$ can be determined. In Figure 7, we show the relations between the ceiling frequencies as functions of $M_{\rm NS}$.

\begin{center}
\begin{figure*}[h]
\label{fig6}
 \includegraphics[width=1.0\columnwidth]{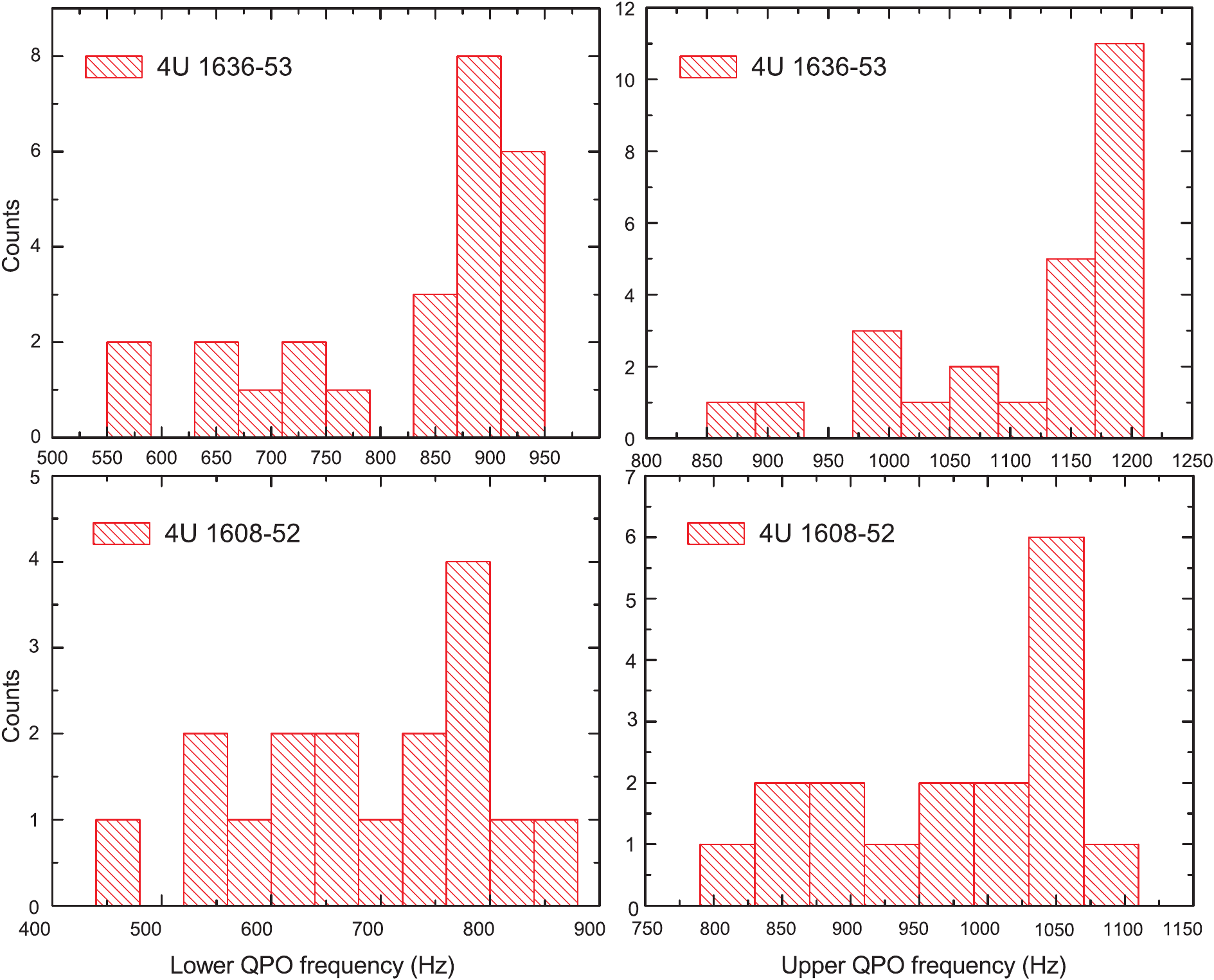}
 \caption{The histograms of the twin kHz QPOs in 4U 1636-53 and 4U 1608-52 (for the measured data: same with the caption of Figure 5). }
\end{figure*}
\end{center}

For 4U 1636--53 we have $\nu_{\rm upper, ceiling}=1181 \pm 13\ {\rm Hz}$ and $\nu_{\rm lower, ceiling}=902 \pm 17\ {\rm Hz}$, corresponding to $M_{\rm NS}=1.90\pm\ 0.01M_{\odot}$ and $1.85\pm\ 0.02M_{\odot}$, respectively. In comparison, $M_{\rm NS}=2.02\ \pm\ 0.12M_\odot$ was obtained by Kaaret et al. (1997) by assuming the maximum maximum $\nu_{\rm upper}=1171\ {\rm Hz}$ as the Kepler frequency at ISCO in a Kerr spacetime.  Similarly, in 4U 1608--52 $M_{\rm NS}=2.03\ \pm\ 0.02M_{\odot}$ or $1.98\ \pm\ 0.03M_{\odot}$ for $\nu_{\rm upper, ceiling}=1042 \ \pm 15\ {\rm Hz}$ or $\nu_{\rm lower, ceiling}=772\ \pm\ 14\ {\rm Hz}$, respectively.

\begin{center}
\begin{figure*}[h]
\label{fig7}
 \includegraphics[width=1.0\columnwidth]{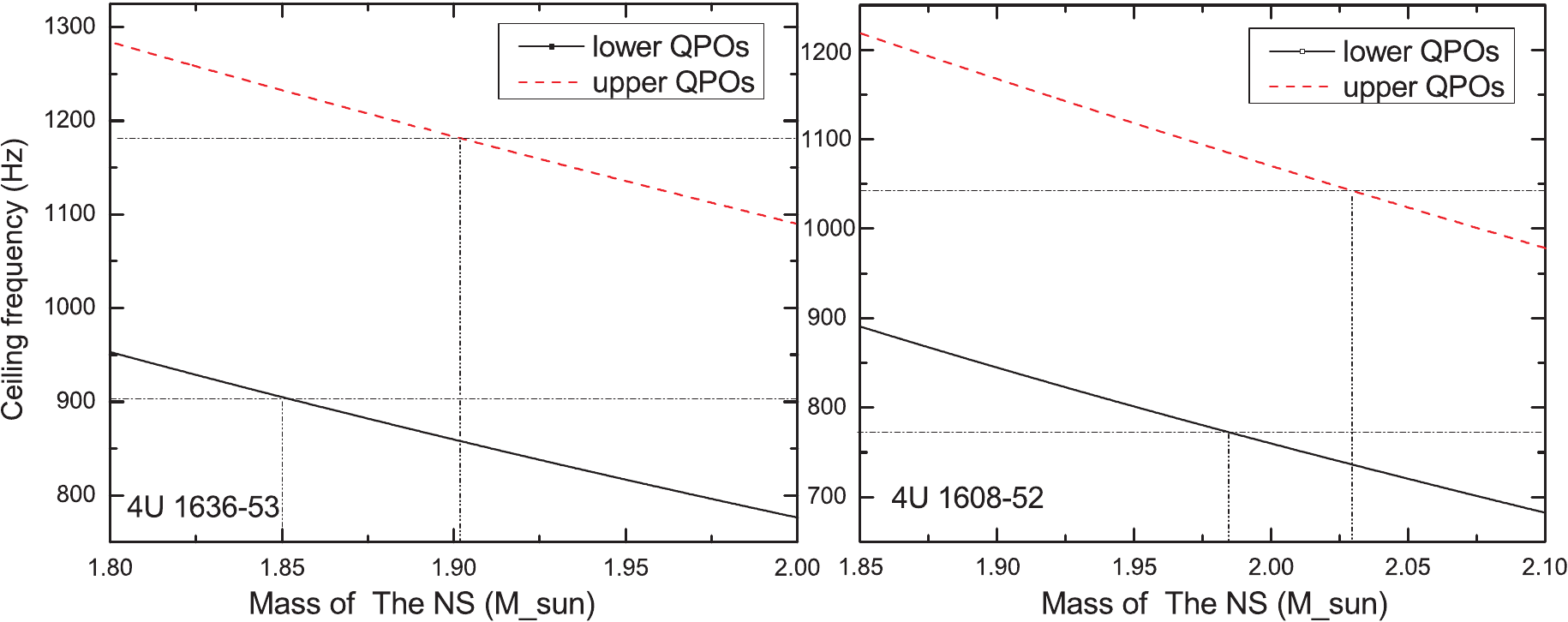}
 \caption{The relation between the ceiling frequency of the twin KHz QPOs and the estimated mass of the NS in 4U 1636-53 and 4U 1608-52. }
\end{figure*}
\end{center}

\subsection{Single kHz QPOs}
The corresponding $\dot{M}$ for single kHz QPOs cannot be identified by the above method for the twin kHz QPOs and can be considered as our model prediction in the right panels of Figure 5. In the left panels of Figure 5, the single numerical solutions in model A are too small to match the observed single kHz QPOs so we will discuss the single kHz QPOs only in model B.

Generally we use a Lorentzian function to describe the finite-width peak in the power spectrum, i.e. QPOs, and the QPOs are confirmed only if the qualify factor is larger than 2. As shown in Figure 8, the single kHz QPOs with low count rate outside the box are related to low qualify factor and the other one of the twin kHz QPOs may also be related to a low qualify factor. The other kHz QPOs with very low qualify factor may be missed and some reported single kHz QPOs may be one of the twin kHz QPOs.

Barret et al. (2006) believed that the origin of $\nu_{\rm upper}$ was different from the one of $\nu_{\rm lower}$ due to the different changing trends of their qualify factors. It is also possible that the qualify factors is related to the different origins of twin kHz QPOs and single kHz QPOs. As shown in the right panel of Figure 8 for 4U 1728--34, three reported ``single" kHz QPOs clustered around the twin kHz QPOs probably belong to twin kHz QPOs, but with one of the twin kHz QPOs missed from detection due to possibly lower signal to noise ratio. On the other hand,  the reported single kHz QPOs in the box of each of the two panels of Figure 8 are located quite distinctively separated from all others, and are thus considered true single kHz QPOs; their frequencies decrease with the increase of the count rate (probably an indicator of accretion rate), as predicted by model B shown in Figure 5.
The real single kHz QPOs in the other sources are identified by the same method as above and the unselected single kHz QPOs listed in the references in the caption of Figure 5 are omitted from Figure 5.

\begin{center}
\begin{figure*}[h]
\label{fig8}
 \includegraphics[width=1.0\columnwidth]{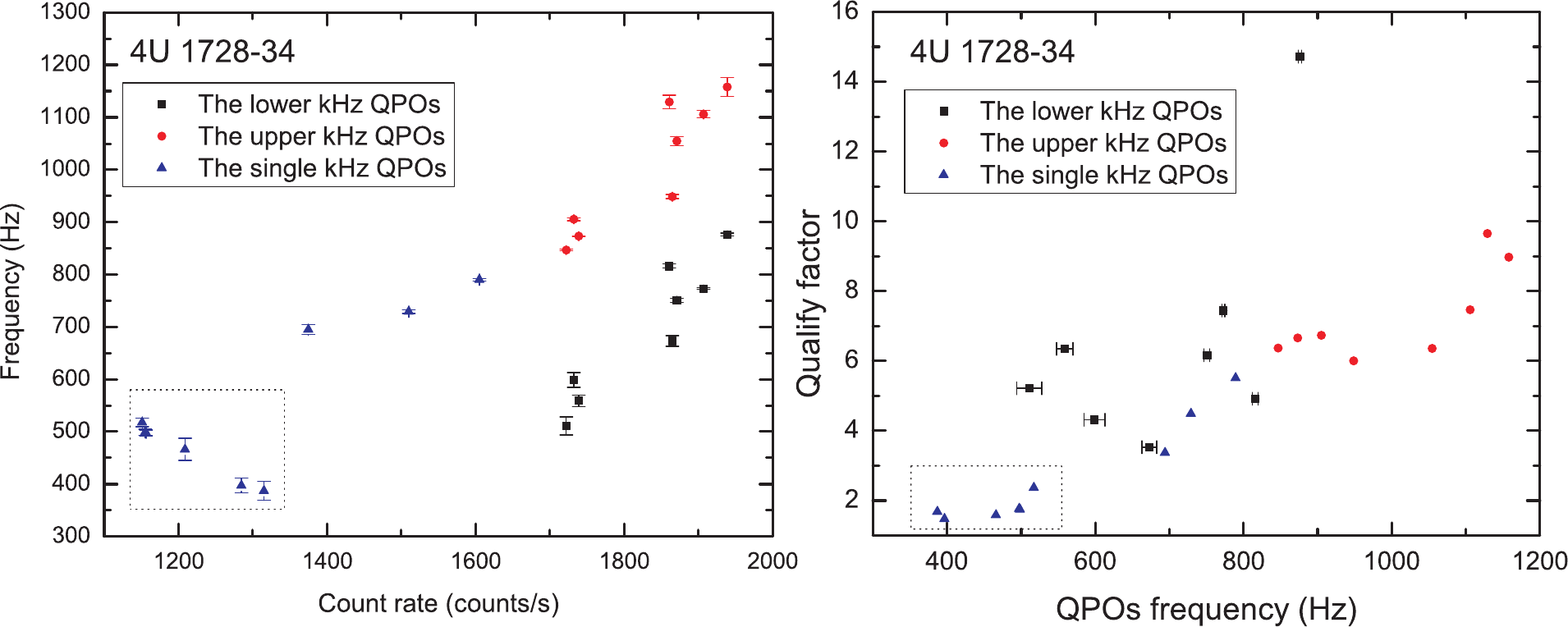}
 \caption{The observation on the count rate of the LMXBs, the frequency of the kHz QPOs and their qualify factor of 4U 1728--34 (for the measured data: Di Salvo et al. 2001). }
\end{figure*}
\end{center}

\section{Discussion}

In this study we only consider that the rotation axis of NSs coincides with the magnetic axis, i.e. the inclination angle is zero. M\'endez et al. (2001) concluded that the properties of the kHz QPOs are determined only by the mass accretion rate through the disk. Lamb et al. (2009) considered that the magnetic inclination is likely to be very small for accretion-powered millisecond pulsars.
Romanova \& Kulkarni (2009) found that the moving spots in the magnetic boundary layer regime might produce QPO features in some cases. The further simulation by Kulkarni \& Romanova (2013) showed that the magnetosphere radius mainly depends on the accretion rate in Equation (24) but not on the misalignment angle of the dipole magnetic field (also see Figure 4 in paper of Kulkarni \& Romanova 2013). It seems that there is little influence on the kHz QPOs from the magnetic inclination because the central frequencies of kHz QPOs in our model are mainly determined by the magnetosphere radius.

There are some complex wave frequency solutions from Eq. (22) and (30) such that the oscillations also grow or decrease in amplitude when the real solutions also exist for the same parameters; however the corresponding periods are too long to be observed. The accretion rates of the other complex wave frequency solutions are too small and the solutions are obtained in unstable accretion region where magnetosphere radius is much larger than the corotation radius and the propeller effect will begin to dominate. We therefore have omitted them and selected only the real solutions, which are considered as the source of the kHz QPOs in NS-LMXBs.
The magnetosphere radius is always regarded as the termination radius
of the accretion onto the NS and it probably determines the position of the boundary layer. Due to the
effect of the centrifugal force, it is often discussed that the magnetosphere radius should be restricted within the corotation radius (Pringle \& Rees 1972; Spruit \& Taam 1993; Rappaport et al. 2004) and it would lead to an accretion process with instability when the magnetosphere radius is more than the corotation radius because of the propeller mechanism. After substituting the compressed magnetosphere radius into our kHz QPOs model, the kHz QPOs can be divided into two parts, i.e., the single kHz QPOs and the twin kHz QPOs in Figure 5. Because the result for the magnetosphere radius from Kulkarni \& Romanova (2013) matches the observation better, we will only discuss the result for model B below. According to the result for the new magnetosphere radius we find that the transition radius ($r_{\rm t}$) is very close to the corotation radius (see Table 1) and so the twin kHz QPOs may originate from the steady accretion process and the single kHz QPOs may mainly originate from the unstable accretion process; the latter may be responsible for the low quality factor of the true single kHz QPOs shown in the box of right panel of Figure 8.

With the decrease of the accretion rate the frequencies of the single kHz QPOs increase outside $r_{\rm t}$ and will exceed the ceiling of the twin kHz QPOs in model B when the accretion rate of LMXBs is very low (about $5*10^{10}\ {\rm g/s}$ for the parameters in the right panel of Figure 3), however the expected high frequency cannot be detected due to the following reasons.
(i) A NS-LMXB in such a very low accretion rate cannot be detected with sufficiently high signal to noise ratio.
(ii) we can confirm kHz QPOs only when the fluctuation of the signal is small enough and signal-noise ratio is high enough.
(iii)  The magnetosphere radius was not simulated by Kulkarni et al. (2013) when the the accretion rate is very low and maybe it can be considered as a uncompressed one, i.e the modes of the kHz QPOs in the compressed magnetic field may convert to the ones in the dipolar magnetic field.

In our model the surface magnetic field of the NS is considered as an invariant and so the accretion rate of the LMXBs that determines the magnetosphere radius is the key parameter. In Figures 3 and 5 the accretion rate is not an observed result because it is determined by our selection from comparing the central frequencies of the observational lower kHz QPOs to our lower numerical solution. Generally the energy spectrum from each observation needs to be fitted to calculate the corresponding physical accretion rate. However the quality of the available data from RXTE is not good enough to do that. This is why we compare $\nu_{\rm upper}$ in the twin kHz QPOs with the observation by means of model derived accretion rate and the detailed relation between the frequencies of kHz QPOs and the measured accretion rate needs to be tested with future observations of better X-ray instruments.

\section{Summary}
In this study we re-examined the MHD model of kHz QPOs and the relation between the frequencies of the kHz QPOs and the accretion rate is predicted {in the two models}. In model A, the magnetic field of a NS keeps its dipolar topology and the accretion disk is compressed due to the magnetic pressure of its dipolar field. In model B, the accretion disk keeps the standard $\alpha$-disk and the magnetosphere of a NS is compressed due to the gas pressure of the standard disk. We find that the results of model B match the observations much better. Our main results are summarized as follows.
\begin{enumerate}
\item The Alfv\'en-like wave and the sound-like wave at the magnetosphere radius only exist under special conditions.
\item We predict that the accretion process of NS-LMXBs may happen in two different areas: sometimes only single kHz QPOs are produced and sometimes twin kHz QPOs are produced. There is a transition radius for the changing kHz QPOs from the single kHz QPOs to the twin kHz QPOs in frequency-accretion rate diagrams, i.e. $r_{\rm t1}$ in model A or $r_{\rm t2}$ in model B as shown in Fig.~3.
\item In model B, the frequency of the single kHz QPOs decreases first and then increases with increasing accretion rate; the transition radius is $r_{\rm t}$, as shown in Fig.~3. All the frequencies of the twin kHz QPOs increase with the increase of the accretion rate.
\item The transition radii ($r_{\rm t}$, $r_{\rm t1}$, $r_{\rm t2}$) are all near the corotation radius.
\item In model B, the lower QPO frequency in a frequency-accretion rate diagram is cut off at low accretion rate and the twin kHz QPOs encounter a top ceiling at high accretion rate due to the restriction of ISCO.
\item In model B, the mass of the NS in 4U 1636--53 is estimated as $1.86-1.90M_\odot$ and the mass of the NS in 4U 1608--52 is estimated as about $1.98-2.03M_\odot$, provided that the observed ceiling frequencies of each NS originate from the magnetosphere radius truncated at the ISCO of the NS.

\end{enumerate}

\acknowledgments

This work is supported by 973 Program of China
under grant 2014CB845802, by the National Natural Science Foundation
of China under grant Nos. 11133002, 11373036, 11203009, 11133001 and 11333004, by the Strategic Priority Research Program "The Emergence of Cosmological Structures" of the Chinese Academy of Sciences, Grant No. XDB09000000, by
the Qianren start-up grant 292012312D1117210, and by the 54th postdoctoral Science Foundation funding of China.


\end{document}